\documentclass[preprint]{elsarticle}
\pdfoutput=1
\usepackage{latexsym, graphicx}
\usepackage{amsmath, amssymb}
\usepackage{color, hyperref}

\begin{document}

\begin{frontmatter}

\title{Quantum mechanical Gaussian wavepackets of single relativistic particles}
\author[a]{Yu-Che Huang}
\ead{ughuang961026@gapp.nthu.edu.tw}
\author[a,b]{Fong-Ming He}
\ead{s112022559@m112.nthu.edu.tw}
\author[b]{Shih-Yuin Lin}
\ead{sylin@cc.ncue.edu.tw}

\affiliation[a]{organization={Department of Physics},
            addressline={National Tsing Hua University}, 
            city={Hsinchu},
            postcode={300044}, 
            country={Taiwan}}
\affiliation[b]{organization={Department of Physics},
            addressline={National Changhua University of Education}, 
            city={Changhua},
            postcode={500207}, 
            country={Taiwan}}

\date{10 September 2023; Corrected 31 July 2026}

\begin{abstract}
We study the evolutions of selected quasi-(1+1) dimensional wavepacket solutions to the Klein-Gordon equation for a relativistic charged particle in uniform motion or accelerated by a uniform electric field in Minkowski space. We explore how good the charge density of a Klein-Gordon wavepacket can be approximately described by a Gaussian state with the single-particle interpretation. We find that the minimal initial width of a wavepacket for a good Gaussian approximation in position space is about the Compton wavelength of the particle divided by its Lorentz factor at the initial moment. This can be considered as a manifestation of relativistic length contraction, which is also observed numerically in the spreading of the wavepacket's charge density. 
\end{abstract}


\end{frontmatter}

\section{Introduction}

In relativistic quantum mechanics, relativistic particles are described by the Klein-Gordon(KG) equation, the Dirac equation, etc., depending on their spins \cite{Gr00}. When the energy density of a particle or experienced by the particle is not too high, while the KG or the Dirac equation may not be reduced to the Schr\"odinger equation in some situations in atomic and nuclear physics,
relativistic quantum mechanics can work nicely with the single-particle interpretation \cite{Gr00}.
At a sufficiently high energy density, however, the KG wavefunctions and the Dirac spinors exhibit some weird behaviors in the viewpoint of non-relativistic quantum mechanics, such as the non-positive-definite charge density and the Klein paradox \cite{Gr00, AG21}. Those behaviors indicate particle-antiparticle production \cite{Gr00, Sc50, Ni70} and thus the breakdown of the single-particle interpretation in quantum mechanics. In those cases, one should consider relativistic quantum field theory \cite{AG21, IZ80} to consistently deal with the many-body nature of the system.

Recently, a linearized effective theory \cite{LH23, Lin23} for single electrons moving in quantum electromagnetic(EM) fields has been constructed in order to address the issue of the Unruh effect \cite{Unr76, DeW79} on the motion of relativistic charged particles \cite{CT99}. That effective theory starts with the total action consisting of (\ref{clAction}) and the one for free EM fields. Then the dynamical variables are quantized canonically by introducing the equal-time commutation relations
\begin{equation}
[ \hat{z}^i, \hat{p}_j ] = i\hbar \delta^i_j \hspace{.5cm} {\rm and} \hspace{.5cm}
[ \hat{A}^\mu_{\bf x}, \hat{\pi}^\nu_{\bf y} ] = i\hbar \eta^{\mu\nu} \delta^3({\bf x}-{\bf y}),
\end{equation}
where $\hat{z}^i$ and $\hat{A}^\mu_{\bf x}$ are the operators for the deviations of the electron position and EM fields from their classical solutions, respectively, and $\hat{p}^i$ and $\hat{\pi}^\mu_{\bf x}$ represent their conjugate momenta.
Such a Dirac quantization implicitly assumes that the single-particle interpretation and the associated Born's rule apply to the (reduced) quantum state of the electrons 
in the Schr\"odinger representation of this effective theory. 

Further, the authors of \cite{LH23,Lin23} considered Gaussian states of single electrons and EM fields. The reasons are obvious:
Gaussian states of quantum systems are mathematically simple yet capable of describing a broad range of physics, including the ground state and squeezed states of harmonic oscillators(HOs), and the vacuum and thermal states of quantum fields. For Gaussian states of two HOs, the degree of quantum entanglement between the HOs is well defined even when the two-HO system is in a mixed state \cite{VW02, LH09}. Taking these advantages, 
a particle-field interacting system in a Gaussian state can be analyzed in details and even non-perturbatively in a long period of evolution if the combined system is linear (e.g. \cite{LH09, LH06, LH07}). 

Nevertheless, it is known that in relativistic quantum mechanics a wavefunction 
that initially looks like a Gaussian function can evolve very differently from the Gaussian wavepackets of the Schr\"odinger equation \cite{RU87, TAL16, GM23}. 
A closed-form example has been provided by Rosenstein and Usher in Ref. \cite{RU87}, where they explicitly demonstrated that the probability density of the Salpeter wavefunction for a particle at rest in (1+1) dimensional Minkowski space with initial width smaller than its Compton wavelength $\lambda^{}_C$ will not concentrate around the classical trajectory after the initial moment, but rather splits into peaks around the lightcone started with the initial position. Then the shape of the probability density of the wavepacket becomes highly non-Gaussian (similar to the behavior of the KG charge density in Figure \ref{WavePacketFree}.) 

This generated the question: In what conditions is a Gaussian probability density of the wavefunctions (in the Schr\"odinger representation of a quantized effective theory such as Refs. \cite{LH23, Lin23}) justified to mimic a nearly Gaussian charge density of the KG wavefunction or Dirac spinor? 
In other words, what is the range of validity or the cutoff for an effective theory like Refs. \cite{LH23, Lin23} for single electrons moving in EM fields? 

To answer the above question about {\it moving} single electrons, merely the wavepackets in Refs. \cite{RU87, DP15} for particles {\it at rest} are not sufficient.  
Thus in this paper, we first generalize the Salpeter wavepacket solution in Ref. \cite{RU87} to the KG wavepacket for relativistic particles in uniform motion, and then construct KG wavepackets initially Gaussian for free particles at constant speed, and for accelerated particles in a uniform electric field. 
We calculate the KG charge densities of our moving KG wavepacket solutions and see whether their shapes look like Gaussian functions in a long period of time. If they do, 
the approximated description using Gaussian wavepackets in the Schr\"odinger representation of an effective theory like \cite{LH23, Lin23}, associated with Gaussian probability densities in the single-particle interpretation, will hopefully be justified. Then we can identify the range of validity the effective theory for moving single electrons after exploring the parameter space.

This paper is organized as follows. In Section \ref{WPinRU87}, we give our wavepacket solutions for free relativistic particles in uniform motion. Our solutions are generalized from the solutions to the Salpeter equation given in \cite{RU87}. Since these solutions also satisfy the KG equation for free particles, we can simply treat them as the KG wavefunctions and calculate the KG charge densities of them. After we study the properties of these KG charge densities, in Section \ref{WPFreeInitGauss} we consider an alternative class of the KG wavefunctions for free particles in uniform motion, which are exactly Gaussian at the initial moment. Similar wavepacket solutions to the KG equation for relativistic charged particles accelerated in a uniform electric field are given and studied in Section \ref{KGuEF}. Combining our results, we find that the minimal initial width for the KG charge density of a Gaussian KG wavefunction evolving like a Gaussian wavepacket centered at its classical trajectory is about $\lambda^{}_C/\gamma^{}_0$, where $\gamma^{}_0 = 1/\sqrt{1-(v^{}_0/c)^2}$ is the Lorentz factor of the particle initially at speed $v^{}_0$. 
Our findings are summarized in Section \ref{summary}. Finally in \ref{ApxPhiScl}, we remark on how the phase of a KG wavepacket solution, which are evaluated around the particle's classical trajectory, evolves in a manner akin to the corresponding classical action.

\section{Wavepackets of free relativistic particles}
\label{FreeRP}

The KG equation for relativistic particles of mass $m$ and charge $q$ moving in EM fields $A^\mu$ reads \cite{Gr00}
\begin{eqnarray}
  \Big[ (\hat{p}_\mu - qA_\mu)(\hat{p}^\mu -qA^\mu) c^2 + m^2 c^4 \Big]\Psi = 0 \label{KGE}
\end{eqnarray}
with the signature $(-,+,+,+)$ and $\hat{p}_\mu = -i\hbar \partial_\mu$.
The KG charge density is defined as
\begin{equation}
  \rho(t,{\bf x}) \equiv {\rm Re}\,\frac{q}{m c^2}\Psi^* (t,{\bf x}) \Big[ i\hbar \partial^{}_t - q A^0_{}(t,{\bf x})\Big] \Psi(t,{\bf x}),
	\label{ChargeDensityDef}
\end{equation} 
which indicates the dominance of anti-particle in the regions where $\rho$ is negative \cite{Gr00}.
If $\rho$ is approximately positive definite and normalizable, then $\rho(t,{\bf x})$ could approximately represent the probability density of finding the particle at $(t,{\bf x})$ 

For free relativistic particles, (\ref{KGE}) reduces to
\begin{eqnarray}
  0= \Big[ \hat{p}_\mu \hat{p}^\mu c^2 + m^2 c^4 \Big]\Psi
	= \left[ (\hbar\partial_t)^2 - \hbar^2 c^2 \nabla^2 + m^2 c^4 \right]\Psi \label{KGFree}
\end{eqnarray}
The solutions of positive energy to (\ref{KGFree}) also satisfies the Salpeter equation or the square-root KG equation \cite{KR11},
\begin{equation}
  i\hbar\partial_t \Psi(t,x) =\left( mc^2\sqrt{1 + \left[\frac{\hbar}{i m c}\right]^2 \nabla_{}^2}\, \right) \Psi(t,x),
	\label{RQMfree}
\end{equation} 
whose form is similar to the Sch\"odinger equation.

Although the probability density in the context of the Salpeter equation is given as $|\Psi(t,x)|^2$, which is positive definite, we will not proceed along this line in this paper. For free particles, while a wavepacket solution $\Psi$ to the Salpeter equation (\ref{RQMfree}) must be a solution to the KG equation (\ref{KGFree}), we will only focus on the well accepted KG charge density $\rho$ of the wavepacket $\Psi$ in (\ref{ChargeDensityDef}). In the presence of background EM fields, the KG equation and the Salpeter equation are not equivalent. In Section \ref{KGuEF}, we will only consider the solutions to the KG equation and the corresponding KG charge density for particles accelerated in a uniform electric field. These KG wavefunctions will not satisfy the Salpeter equation with the same background fields.

\subsection{A wavepacket in uniform motion}
\label{WPinRU87}

A wave-packet solution to (\ref{RQMfree}) and so (\ref{KGFree}) has been given by Rosenstein and Usher in Ref. \cite{RU87}. Following the same method, we generalize their result to the case of free particles in uniform motion below. Starting with the ansatz (independent of $y$ and $z$),
\begin{equation}
  u^{}_p(t,x) = \exp \left[ -\frac{1}{\hbar}(\vartheta +it)W(p) + \frac{i}{\hbar}
	p(x-v^{}_0 t) \right], \label{RPfreeAns}
\end{equation}
where $v^{}_0$ 
is the velocity of the classical trajectory of the particle in the $x$-direction, and $c\vartheta$ corresponds to the initial width of the probability density $|\Psi|^2$ of the wavepacket \cite{RU87} in the case of $v^{}_0=0$.
Inserting the above ansatz into (\ref{RQMfree}), we obtain
\begin{equation}
  W(p) = mc^2 \sqrt{1+\left(\frac{p}{mc} \right)^2} - p v^{}_0, 
\end{equation}
which is positive for all $p=m v \gamma$ since $v^{}_0, v < c$.
When $p=p^{}_0=m v^{}_0 \gamma^{}_0$ with $\gamma^{}_0 \equiv 1/\sqrt{1-(v^{}_0/c)^2}$, one has $W(p^{}_0) = -L_{cl}$ in value, where the classical Lagrangian $L_{cl} = -mc^2\gamma^{}_0$ can be read off from (\ref{clActionFree}). 

Taking the superposition of $u^{}_p(t,x)$, 
\begin{equation}
  \Psi(t,x) = {\cal N} \int dp\, u_p(t,x) \tilde{\psi}(p) \label{IntFormPsi}
\end{equation}
with the simplest choice, $\tilde{\psi}(p)=1$ for all $p$, we get 
\begin{eqnarray}
  \Psi(t,x) = {\cal N} \int dk \exp \left[ -\frac{mc^2}{\hbar} (\vartheta+it)\sqrt{1+k^2} +\frac{mc}{\hbar}(\vartheta v^{}_0+i x) k \right] \label{PsiRelFreeIntk}
\end{eqnarray}
where $k \equiv p/(mc)$. 
Obviously the above $\Psi(t,x)$ is also a solution to the KG equation (\ref{KGFree}). 
Let $k = \sinh \kappa'$ and integrate over $\kappa'$, we obtain the closed form of the wave-packets as
\begin{equation}
 \Psi(t,x) =\sqrt{\frac{mc}{\hbar\pi\gamma^{}_0 K_1\left(\frac{2 mc^2 \vartheta}{\hbar\gamma^{}_0}\right)}}\,\frac{(\vartheta+ i t)c}{F(t,x)}
  K_1\left(\frac{mc}{\hbar}F(t,x)\right), \label{PsiRelFree}
\end{equation}
where $K_1(z)\equiv\int_0^\infty d\kappa \, e^{-z \cosh \kappa} \cosh \kappa$ is the modified Bessel function of the second kind, and 
$F(t,x)\equiv \sqrt{\left(x-i v^{}_0  \vartheta \right)^2 -c^2 \left(t-i\vartheta\right)^2}$. 
The normalization factor ${\cal N}$ is obtained by requiring $\int_{-\infty}^\infty dx |\Psi|^2 =1$ 
in the context of the Salpeter equation, 
most conveniently using the wavefunction $\Psi$ in the form of (\ref{IntFormPsi}). Similar integration gives $\langle x \rangle =\int_{-\infty}^\infty dx\, x |\Psi|^2 = v^{}_0 t$. 
Note that the dependence of $c\vartheta$ on the initial width of $|\Psi|^2$ is not linear, 
though $c\vartheta$ is a monotonic increasing function of the initial width in the parameter range we explored.

\subsubsection{Initial width and non-Gaussianity}

\begin{figure} 
\includegraphics[width=3.9cm]{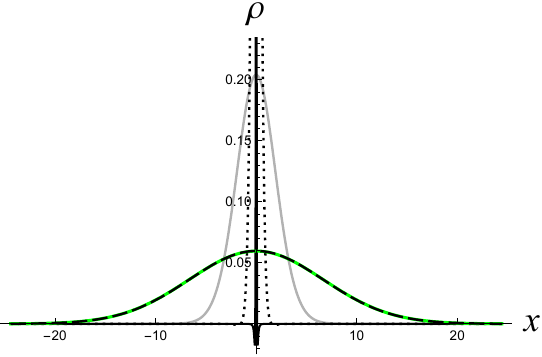}
\includegraphics[width=3.9cm]{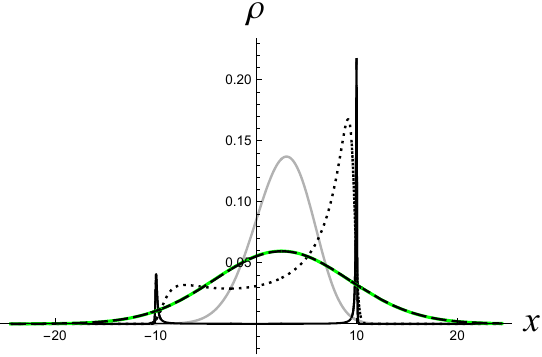}
\includegraphics[width=3.9cm]{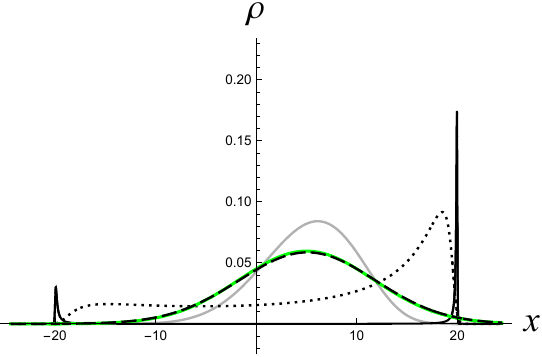}
\caption{The (scaled) spatial distribution of the KG charge density $\rho(t,x)$ (defined in (\ref{ChargeDensityDef}) and normalized by $\int_{-\infty}^\infty dx\rho =1$ here)  for $t=0$ (left), $10$ (middle), and $20$ (right) with various initial widths of wavefunction (\ref{PsiRelFree}). Here $c=\hbar = m \equiv 1$, and so $x$ is in the unit of the reduced Compton wavelength of the particle $\bar{\lambda}^{}_C = \hbar/(mc) = \lambda^{}_C/(2\pi)$. 
The black dashed, gray, black dotted, and black solid curves in each plot represent the charge densities $\rho$ of wavefunction (\ref{PsiRelFree}) with $c\vartheta=100$, $10$, $1$, and $0.1$, all moving at a constant speed $v^{}_0 = c/4$ to the right. 
Note that the black solid curves have been scaled down to $1/20$ of their original values. The green curves represent the Gaussian function $\rho^{}_G(t,x) = (\sigma^{}_0\sqrt{\pi})^{-1} e^{-(x-v_0 t)^2/\sigma_0^2}$, where we choose $\sigma^{}_0 = 9.468$ 
(constant in time) as a reference for the black dashed curve ($c\vartheta=100$).  At $t=20$ the spreading of the black dashed curve is not significant yet.} 
\label{WavePacketFree}
\end{figure}

Let us consider $\Psi$ in (\ref{PsiRelFree}) as a KG wavefunction satisfying (\ref{KGFree}) for a relativistic free particle, and insert it into (\ref{ChargeDensityDef}) to calculate the KG charge density.
As shown in Figure \ref{WavePacketFree}, the KG charge density $\rho$ of wavepacket solution (\ref{PsiRelFree})  behaves like a Gaussian function peaked around the classical trajectory if the value of parameter $c\vartheta$ ($\gamma$ in \cite{RU87}) is sufficiently large (e.g., the dashed curve with $c\vartheta=100$ in Figure \ref{WavePacketFree}.) 
This is consistent with the observation on $|\Psi|^2$ in Ref. \cite{RU87} in the context of the Salpeter equation.
Note that in Figure \ref{WavePacketFree}, the Lorentz factor of the particle is $\gamma^{}_0 \approx 1.03$, and so the particle there is not in a highly relativistic motion.

If the initial width is well below the Compton wavelength of the particle [e.g. the case of $c\vartheta = 0.1$ in Figure \ref{WavePacketFree}, corresponding to the width $2\sigma = 0.092 \bar{\lambda}^{}_C \ll 2\pi \bar{\lambda}^{}_C = \lambda^{}_C$ at $t=0$ in Figure \ref{WavePacketFreeRUSimGauss} (right)], the charge density behaves very differently from a moving Gaussian function. At $t=0$, while the shape of the Salpeter probability density $|\Psi|^2$ is still close to a Gaussian function and the wavefunction $\Psi$ here should include only the positive-energy modes as they are solutions to (\ref{RQMfree}), the shape of the corresponding KG charge density $\rho$ has been non-Gaussian: There exists regions where the charge density is negative, indicating the presence of antiparticles [black solid curve in Figure \ref{WavePacketFree} (left)]. So the single-particle interpretation fails here, 
and $\rho(t,{\bf x})$ cannot be interpreted as a probability density of finding the particle at $(t,{\bf x})$. Furthermore, when $t>0$, the charge density $\rho$ splits into two peaks around the left and right edges of the lightcone in the $t$-$x$ diagram [Figure \ref{WavePacketFree} (middle) and (right)].  Obviously, such kind of double-peak distributions is highly non-Gaussian.

If we take the values of $c\vartheta$ from $100$ down to $0.1$ continuously, in Figure \ref{WavePacketFree} one can see that the shape of the charge density distribution changes continuously from almost Gaussian to highly non-Gaussian functions. Actually, the gray curves of $c\vartheta = 10$ in Figure \ref{WavePacketFree} are quite close to Gaussian functions, though a small asymmetry about the classical particle position can be seen.

\subsubsection{Similarity to Gaussian functions}
\label{SimGauss}

\begin{figure} 
\includegraphics[width=5.7cm]{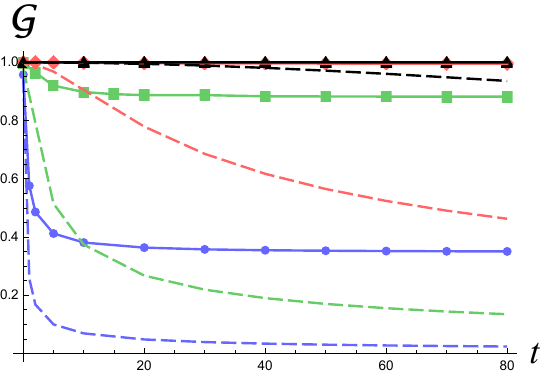}\hspace{.5cm}
\includegraphics[width=5.7cm]{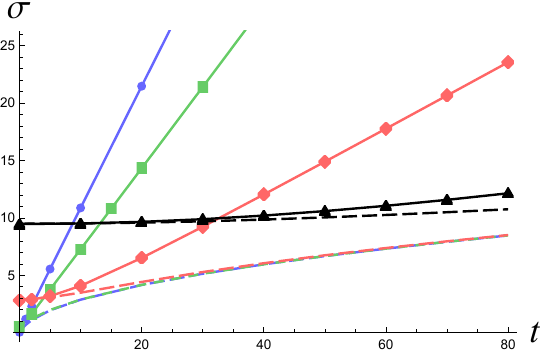}
\caption{(Left) Time evolution of ${\cal G}^{}_\Psi$ [dashed curves, see Eq. (\ref{GauPsi})] and ${\cal G}^{}_\rho$ [solid curves, see Eq. (\ref{GauRho})] of the wavepackets considered in Figure \ref{WavePacketFree}. The black (triangle), red (diamond), green (square), and blue (circle) curves represent the cases of $c\vartheta = 100$, $10$, $1$, and $0.1$, respectively. (Right) The half-widths $\sigma$ producing the values of ${\cal G}^{}_\Psi$ (dashed curves) and ${\cal G}^{}_\rho$ (solid) in the left plot. At $t=0$, the initial widths of the best-fit Gaussian functions to the charge densities $\rho$ for wavefunctions (\ref{PsiRelFree}) with $c\vartheta = 100$, $10$, $1$, and $0.1$ are $2\sigma = 18.94$, $5.654$, $1.048$, and $0.092$, respectively.}
\label{WavePacketFreeRUSimGauss}
\end{figure}

To see how close the above wavefunctions to a moving Gaussian wavepacket of half-width $\sigma$ and momentum $\bar{p}(t)$ centered at $x=\bar{x}(t)$, namely,
\begin{equation}
  \varphi^{}_G(t,x) \equiv \frac{1}{\sqrt{\sigma\sqrt{\pi}}}\exp 
	\left[ -\frac{1}{2\sigma^2}\big(x - \bar{x}(t)\big)^2 +\frac{i}{\hbar} \bar{p}(t) x \right]
\end{equation}
with $\bar{x}(t)=\bar{v} t$ and $\bar{p}(t) = m \bar{v} \bar{\gamma} = m \bar{v}/\sqrt{1-(\bar{v}/c)^2}$ for a free particle at constant velocity $\bar{v}$ in the $x$-direction, we calculate the projection
\begin{equation}
  {\cal G}^{}_\Psi = \max_{\sigma} \left| \int dx \varphi^*_G(t,x)\Psi(t,x)\right|^2 \label{GauPsi}
\end{equation}
with the best fit of $\sigma$ producing the maximum value of ${\cal G}^{}_\Psi$. The similarity of the KG charge densities $\rho$ of the above wavefunctions to a normalized Gaussian function of the half-width $\sigma$ centered at $x=\bar{x}(t)$,
\begin{equation}
  \rho^{}_G(t,x) \equiv \left|\varphi^{}_G(t,x)\right|^2 = 
	\frac{1}{\sigma\sqrt{\pi}}\exp \left[ -\frac{1}{\sigma^2}\big(x - \bar{x}(t)\big)^2 \right],
\end{equation}
may also be estimated by calculating
\begin{equation}
  {\cal G}^{}_{\rho}(t) = \max_{\sigma} \frac{\int dx \, \sqrt{\rho^{}_G (t,x) \rho(t,x)}}{\sqrt{\int dx \, |\rho(t,x)|}}. \label{GauRho}
\end{equation}
Once $\rho(t,x)$ has negative regions in $x$, the square root in the integrand of the above numerator will generate complex values, though their imaginary parts are all negligible in Figures \ref{WavePacketFreeRUSimGauss}, \ref{GaussWavepacketFreeSimGauss}, and \ref{WavePacketinESimGauss}. 

In Figure \ref{WavePacketFreeRUSimGauss} (left), we show the quantities ${\cal G}^{}_\Psi$ (dashed curves) and ${\cal G}^{}_\rho$ (solid curves) for those wavepackets considered in Figure \ref{WavePacketFree}. One can see that the larger value of $c\vartheta$ [corresponding to the larger initial width $2\sigma$ in Figure \ref{WavePacketFreeRUSimGauss} (right)] is, the lower decay rate of the similarity to Gaussian functions ${\cal G}^{}_\Psi$ or ${\cal G}^{}_\rho$ will be. 
For the same wavepacket solution, ${\cal G}^{}_\Psi$ drops faster than ${\cal G}^{}_\rho$ as $t$ increases. This is because the phase of $\Psi$ becomes highly nonlinear in $x$ for $t>0$ and evolves very quickly in a complicated way in $t$ such that the projection in (\ref{GauPsi}) get worse very quickly, while ${\cal G}^{}_\rho$ is not sensitive to the phase.

The corresponding half-widths $\sigma$ for $\Psi$ (dashed curves) and $\rho$ (solid curves) are shown in Figure \ref{WavePacketFreeRUSimGauss} (right), where one can see that the best fit of the half-width $\sigma$ for wavefunction $\Psi$ tends to overlap as $t\to\infty$. In contrast, for each wavepacket solution the half-width $\sigma$ of the KG charge density $\rho$ grows linearly in time for sufficiently large $t$ with a rate depending on the parameter $c\vartheta$. The larger $c\vartheta$ is, the larger initial width $2\sigma$ and the lower spreading rate of $\rho$ will be. When the initial width $2\sigma$ is well below the Compton wavelength, the spreading rate can exceed the speed of light (blue solid curve) as the shape of $\rho$ gets highly non-Gaussian in Figure \ref{WavePacketFree}.

\subsubsection{Momentum spectrum}
\label{MtmSpec}

\begin{figure} 
\includegraphics[width=5.7cm]{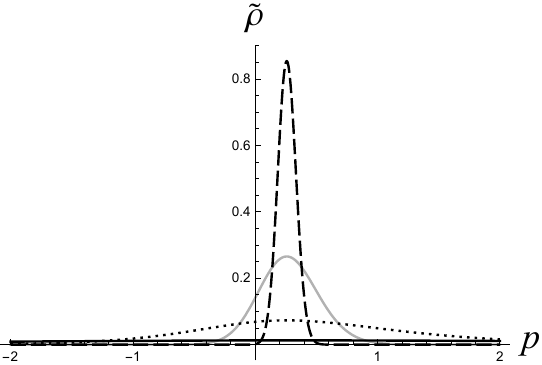}\hspace{.5cm}
\includegraphics[width=5.7cm]{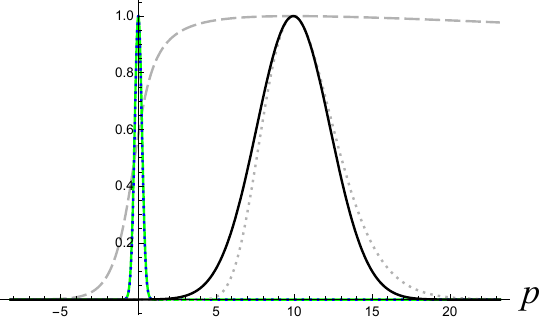}
\caption{(Left) The black dashed, gray, black dotted, and black solid curves represent $\tilde{\rho}(t,p)$ in (\ref{rhotpFree}) for the wavepackets with the parameter $c\vartheta=100$, $10$, $1$, and $0.1$, respectively, for a free particle with $v^{}_0=c/4$ ($\gamma^{}_0\approx 1.03$). Other parameters have the same values as those in Figure \ref{WavePacketFree}.\,\,\,
(Right) Comparison of the Gaussian distribution $|\tilde{\psi}(p)|^2$ from (\ref{psipGaussian}) with $\sigma^{}_0=0.3$ (black) and the distribution (\ref{rhotpFree}) with $c\vartheta= 0.3$ (gray dashed) and $c\vartheta=96$ (gray dotted) for a particle in relativistic motion with $\gamma^{}_0 =10$. Here the distribution functions are normalized to $1$ at $p=p^{}_0 \approx 9.95$ ($m=c=\hbar=1$).
The gray dotted curve [(\ref{rhotpFree}) with $c\vartheta=96$] is quite close to the black curve [(\ref{psipGaussian}) with $\sigma^{}_0=0.3$], so their profile in position space at the initial moment would be similar. One can see that in this regime the value of $c\vartheta$ in (\ref{rhotpFree}) can be very different from the initial width of the corresponding KG charge density in position space (about $2\sigma^{}_0 = 0.6$).
We also compare the absolute square of (\ref{psipGaussian}) with $\sigma^{}_0= 3$ (green) and (\ref{rhotpFree}) with $c\vartheta= 10$ (blue dotted) for $v^{}_0=0$. These two curves almost overlap, and the tails in the region $|p| > 1$ are negligible.
}
\label{WavePacketFreeP}
\end{figure}

The peak-splitting of the charge density with initial width well below its Compton wavelength indicates that the momentum spectrum of the corresponding wavepacket is so broad that many modes of high positive and negative momenta are included. The speed of those high-momentum modes are all close to the speed of light, and so they ``pile up" or constructively interfere around the lightcone in the $t$-$x$ diagram in Figure \ref{WavePacketFree}. To see this more clearly, let
\begin{equation}
  \tilde{\Psi}(t,p) = \int \frac{dx}{\sqrt{2\pi\hbar}} e^{-\frac{i}{\hbar} p x} \Psi(t,x) = {\cal N} \sqrt{2\pi \hbar}\, u^{}_p(t,0)\textsl{}
\end{equation}
with $u^{}_p(t,x)$ given in (\ref{RPfreeAns}).
The momentum spectrum of $\Psi(t,x)$ can be observed via the distribution function 
\begin{equation}
  \tilde{\rho}(t,p) \equiv \left|\tilde{\Psi}(t,p)\right|^2 = 2\pi \hbar |{\cal N}|^2 u_p^*(t,0) u^{}_p(t,0) =  
	2\pi\hbar |{\cal N}|^2 e^{-\frac{2}{\hbar}\vartheta W(p)}, 	\label{rhotpFree}
\end{equation}
which has the maximum at $p = p^{}_0 = mv^{}_0\gamma^{}_0$ since $W(p)$ has the minimum there ($W'(p^{}_0)=0$).  
As shown in Figure \ref{WavePacketFreeP} (left), the smaller value of $c\vartheta$ is, the wider range of modes with $p$ around $p^{}_0$ will contribute. When $c\vartheta < O(\bar{\lambda}^{}_C)$, there will be many modes of $|p| > mc = \hbar/\bar{\lambda}^{}_C$ involved in the wavepacket. These modes are moving at speeds close to $c$ and would constructively interfere around the lightcone. 
In particular, the momentum spectra of the cases with $c\vartheta=1$ (black dotted curve) and $0.1$ (black solid) are significantly nonzero in the region of $p < -mc = -1$. These modes form the significant left-moving peaks apart from the main peaks of the charge densities with $c\vartheta = 1$ and $0.1$ in Figure \ref{WavePacketFree}.

\subsection{Wavefunctions initially Gaussian}
\label{WPFreeInitGauss}

In Figures \ref{WavePacketFree} and \ref{WavePacketFreeRUSimGauss} we have seen that the minimal initial width that the charge density of wavefunction (\ref{PsiRelFree}) can be approximated as a Gaussian function is roughly $5.65 \times$ to $18.94\times$ $\lambda^{}_C/(2\pi)\approx 1 \times$ to $3\times$ $\lambda^{}_C$, corresponding to $c\vartheta = 10$ to $100$. This is not a definite lower limit, anyway. In this section we will show that the minimal initial width for a good Gaussian approximation to another class of slowly moving wavepackets can also reach the Compton wavelength $\lambda^{}_C$. If the wavepacket in that class are in highly relativistic motion, the minimal initial width for Gaussian approximation can be even smaller in the rest frame.

Starting with (\ref{IntFormPsi}), one can choose $\vartheta=0$ and set $\tilde{\psi}(p)=\exp [-\frac{1}{\hbar}\tilde{\vartheta} W(p) ]$. Then $u_p(t,x)|^{}_{\vartheta=0} = \tilde{u}_p(t,x) \equiv \exp \left\{ -\frac{i}{\hbar} \left[ E(p) t - p x\right] \right\}$ with $E(p) = \sqrt{m^2 c^4 + p^2 c^2}$ becomes the conventional plane-wave solution while one still obtains $\Psi(t,x)$ in (\ref{PsiRelFree}) with $\vartheta$ replaced by $\tilde{\vartheta}$. Thus, we could liberate ourselves from (\ref{PsiRelFree}) by choosing $\vartheta=0$ and introducing an alternative $\tilde{\psi}(p)$, which may be designed to make the wavepacket acting more ``classically". 

Suppose the initial KG wavefunction at $t=0$ is exactly a Gaussian function of half-width $\sigma^{}_0 > 0$ and momentum $p^{}_0$, 
centered at $x=x^{}_0$,
\begin{equation} 
  \Psi(0,x) = \frac{1}{\sqrt{\sigma^{}_0\sqrt{\pi}}} \exp \left[ -\frac{(x-x^{}_0)^2}{2\sigma_0^2}+\frac{i}{\hbar}p^{}_0 (x-x^{}_0)\right].
	\label{Psi0Gauss}
\end{equation}
The wavefunction for $t\ge 0$ can be constructed using the superposition of $\tilde{u}_p(t,x)$ with the coefficient
\begin{equation}
  \tilde{\psi}(p) = \int \frac{dx}{2\pi \hbar} e^{-\frac{i}{\hbar}p x} \Psi(0,x) = \frac{\sqrt{\sigma^{}_0}}{\hbar\sqrt{2 \pi^{\frac{3}{2}}}}
	\exp \left[ -\frac{\sigma_0^2}{2\hbar^2} (p-p^{}_0)^2 - \frac{i}{\hbar}p x^{}_0\right], \label{psipGaussian}
\end{equation}
which gives 
\begin{eqnarray}
  \Psi(t,x) &=& \int dp\, \tilde{u}_p(t,x)\tilde{\psi}(p) \nonumber\\
	&=& \frac{\sqrt{\sigma^{}_0}}{\hbar\sqrt{2 \pi^{\frac{3}{2}}}}\int dp\, e^{-\frac{i}{\hbar}
	\left[ E(p)t -p(x-x^{}_0)\right] -\frac{\sigma_0^2}{2\hbar^2} (p-p^{}_0)^2}.
	\label{GaussianPsiFree}
\end{eqnarray}

\begin{figure} 
\includegraphics[width=5.7cm]{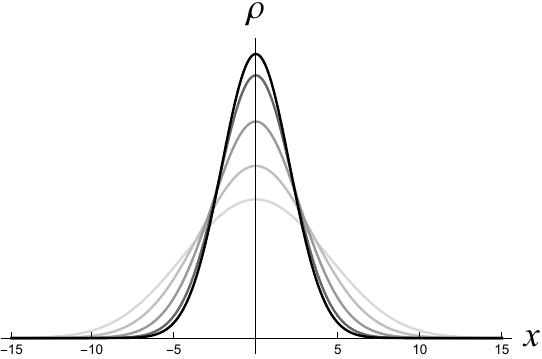}\hspace{.5cm}
\includegraphics[width=5.7cm]{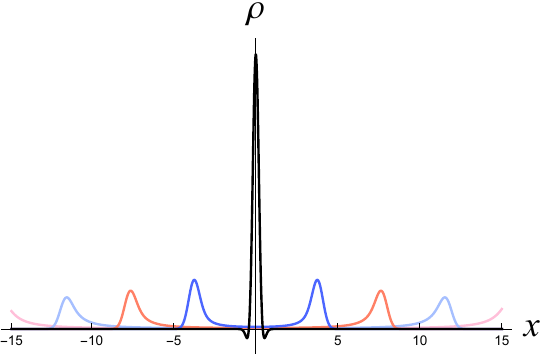}\\
\includegraphics[width=5.7cm]{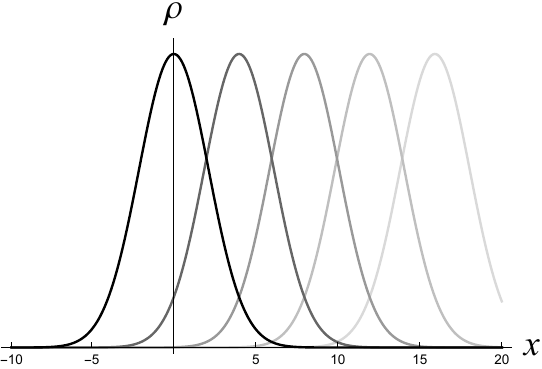}\hspace{.5cm}
\includegraphics[width=5.7cm]{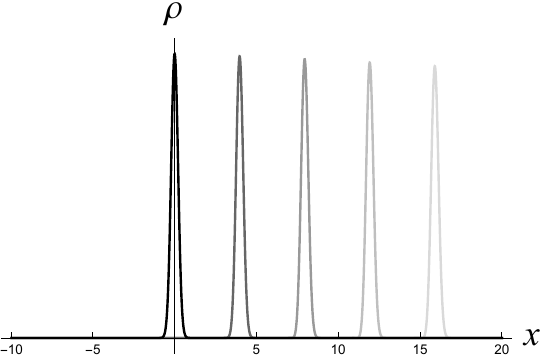}
\caption{Time evolution of the KG charge density $\rho(t,x)$ of wavepacket (\ref{GaussianPsiFree}) with the parameter values $x^{}_0=0$ and $(\sigma^{}_0, \gamma^{}_0) = (3, 1)$ (top-left), $(3, 10)$ (lower-left), $(0.3, 10)$ (lower-right), and $(0.3, 1)$ (upper-right). The curves from dark to light represent $\rho$ at $t=0, 4, 8, 12$, and $16$.}
\label{GaussWavepacketFree}
\end{figure}

\begin{figure}
\includegraphics[width=5.7cm]{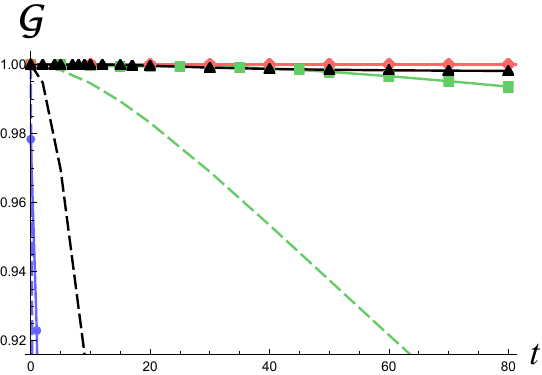}\hspace{.5cm} 
\includegraphics[width=5.7cm]{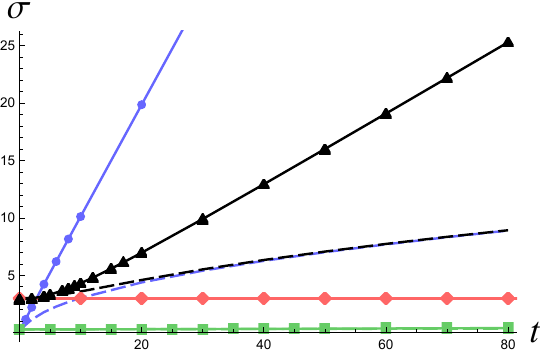}
\caption{(Left) Time evolution of ${\cal G}^{}_\rho$ (solid curve) and ${\cal G}^{}_\Psi$ (dashed) of the wavepackets considered in Figure \ref{GaussWavepacketFree}. The red solid line and the red dashed line cannot be distinguished in this plot. (Right) Time evolution of the half-width $\sigma$ producing the values of ${\cal G}^{}_\rho$ (solid curves) and ${\cal G}^{}_\Psi$ (dashed) in the left plot. The red (green) solid line and the red (green) dashed line are not distinguishable in this plot. Here the black (triangle), red (diamond), green (square), and blue (circle) curves represent the cases of $(\sigma^{}_0, \gamma^{}_0)=(3,1)$, $(3,10)$, $(0.3, 10)$, and $(0.3,1)$, respectively.}
\label{GaussWavepacketFreeSimGauss}
\end{figure}

Four examples are given in Figure \ref{GaussWavepacketFree}.
In the left column, one can see that the KG charge density $\rho$ of the wavepackets of initial width $2\sigma^{}_0 = 6$ in the unit of the reduced Compton wavelength $\lambda^{}_C/(2\pi)$ (namely, $2\sigma^{}_0 = 6 \lambda^{}_C/(2\pi) \approx \lambda^{}_C$) are still very close to Gaussian functions, even in the case of $p^{}_0=0$ (upper-left). Indeed, the momentum spectrum of the wavefunction (\ref{GaussianPsiFree}) with $\sigma^{}_0 = 3$ [the green curve in Figure \ref{WavePacketFreeP} (right)] is mostly restricted in the interval $|p|< mc = 1$.

With initial width well below the Compton wavelength, in contrast, 
while the wavefunction (\ref{GaussianPsiFree}) of a free particle at rest ($p^{}_0=0$ and $\gamma^{}_0=1$) is exactly Gaussian at the initial moment $t=0$, the KG charge density $\rho$ of the initial wavefunction can be very non-Gaussian. 
In the upper-right plot of Figure \ref{GaussWavepacketFree} ($\sigma^{}_0 = 0.3$ and $\gamma^{}_0=1$), one can see that $\rho$ at $t=0$ has negative regions where the antiparticle density dominates, indicating that the single-particle interpretation fails in this case.
In the same plot, $\rho(t,x)$ also splits into two peaks moving apart from each other as $t$ increases, as its wide momentum distribution includes significant contributions from the modes of $|p| > mc$. 
These non-Gaussian features are similar to the case of $c\vartheta=0.1$ in Figure \ref{WavePacketFree} where the wavepackets have $p^{}_0/m = v^{}_0 \gamma^{}_0 < 1$ (non-relativistic motion). 

For the initial momentum $p^{}_0 \gg mc$, we observed that the left-moving peak moving apart from the main peak will be well suppressed if the Gaussian distribution in momentum space has a half-width $\hbar/\sigma^{}_0$ less than about $p^{}_0/3$
[the black curve in Figure \ref{WavePacketFreeP} (right)],
namely, $|\tilde{\psi}(-1)|^2/|\tilde{\psi}(p^{}_0)|^2 \approx e^{-\sigma_0^2 p_0^2/\hbar^2} < e^{-9} \approx 10^{-4}$ with $\tilde{\psi}(p)$ in (\ref{psipGaussian}). Thus, to get a good Gaussian approximation to a charge density with only one peak (around $x = v^{}_0 t$), the minimal initial half-width $\sigma^{}_0 \approx 3\hbar/p^{}_0\approx 3\hbar/(m c \gamma^{}_0) = 3\lambda^{}_C/(2\pi \gamma^{}_0)$ in position space can be very small in the direction of motion for a highly relativistic particle ($\gamma^{}_0 \gg 1$) [see Figures \ref{GaussWavepacketFree} (lower-right) and Figure \ref{WavePacketFreeP} (right).]  Indeed, with the same initial width below the Compton wavelength, the non-Gaussianity in Figure \ref{GaussWavepacketFree} (upper-right) for the case of $(\sigma^{}_0, \gamma^{}_0)=(0.3,1)$ is suppressed in Figure \ref{GaussWavepacketFree} (lower-right) for $(\sigma^{}_0, \gamma^{}_0)=(0.3,10)$.  
The Gaussian approximation for the KG charge density in Figure \ref{GaussWavepacketFree} (lower-right) is as good as the one in Figure \ref{GaussWavepacketFree} (upper-left), and $\sigma^{}_0 = 3/\gamma^{}_0$ in both cases [also compare the green and black curves in Figure \ref{GaussWavepacketFreeSimGauss} (left).] 

Combining the above observations, we find that the Gaussian approximation for the KG charge density $\rho$ of wavefunction (\ref{GaussianPsiFree}) would be good if the initial-width parameter of the wavepacket $2\sigma^{}_0$ is well above $\lambda^{}_C/\gamma^{}_0$. The presence of $\gamma^{}_0$ may be considered as a manifestation of length contraction.

Compare the upper and the lower plots in the same column in Figure \ref{GaussWavepacketFree}, one can see that for the same initial half-width $\sigma^{}_0$, the similarity of $\rho$ to Gaussian functions drops slower for a faster particle in the laboratory frame. This can be confirmed by comparing the black and red curves, and comparing the blue and green curves in Figure \ref{GaussWavepacketFreeSimGauss} (left).
Moreover, for sufficiently large $t$, the half-width $\sigma$ of the best-fit Gaussian function to the charge density $\rho$ grows linearly in $t$, as shown in Figure \ref{GaussWavepacketFreeSimGauss} (right), and the spreading rate in the highly non-Gaussian case (blue curve) can exceed the speed of light. With the same initial half-width $\sigma^{}_0$, the spreading rate of $\rho$ is lower for a particle moving faster.
For particles in uniform motion, however, one could not tell such a dependence of the spreading rate on the particle speed is from time dilation, and/or from length contraction. This will become clearer in the case of accelerated particles.

\section{KG wavepacket of particles in uniform electric field in Minkowski coordinates}
\label{KGuEF}

In shaping the initial wavefunction at $t=0$ for a charged particle in a uniform electric field, it is convenient to choose the EM four-potential $A^\mu = (0, -{\cal E} t,0,0)$, which gives the electric field ${\bf E} = -c F_{0j} \hat{x}^j = {\cal E} \hat{x}^1$. The corresponding KG equation reads ($c=\hbar=1$, ${\bf x}\equiv (x,y,z)$)
\begin{equation}
  \left[ \partial_t^2 + (-i\partial_x + q {\cal E} t)^2 - \partial_y^2-\partial_z^2 +m^2\right] \Psi(t,{\bf x}) = 0.
\end{equation}
Inserting the ansatz $\Psi = e^{i {\bf p}\cdot{\bf x}} \psi^{}_{\bf p}(t)$ into the above equation, one has 
\begin{equation}
  \left[ \partial_t^2 + (p^{}_x + q {\cal E} t)^2 + M^2 \right]\psi^{}_{\bf p}(t) = 0
\end{equation}
with $M^2 = m^2 + p_y^2 + p_z^2$. Below we take $p^{}_y = p^{}_z=0$ for simplicity. 
The general solution to the above equation is \cite{Ni70}
\begin{equation}
  \psi^{}_{\bf p}(t) = c^+_{\bf p} D^{}_{-\frac{1}{2}-\frac{i M^2}{2 F}}\left[\frac{i+1}{\sqrt{F}}(p^{}_x + Ft)\right]
	+ c^-_{\bf p} D^{}_{-\frac{1}{2}+\frac{i M^2}{2 F}}\left[\frac{i-1}{\sqrt{F}}(p^{}_x + Ft)\right], \label{psiDfn}
\end{equation}
where $D^{}_\nu (z)$ are the parabolic cylinder functions, $F \equiv q{\cal E}$, and $c^\pm_{\bf p}$ are constants of time.
To illustrate a wavefunction initially identical to Eq. (\ref{Psi0Gauss}) at $t=0$, we choose
\begin{equation}
  \Psi(t,{\bf x}) = \int d^3 p\,  e^{i {\bf p}\cdot{\bf x}} \psi^{}_{\bf p}(t) \label{PsiUAC}
\end{equation}
with the constants of time
\begin{equation}
c^{\pm}_{\bf p}
={\cal N}^{}_{p^{}_x}\sqrt{2\sigma^{}_0\sqrt{\pi}}\,\delta(p^{}_y)\delta(p^{}_z) D_{-\frac{1}{2}\pm 
  \frac{iM^2}{2F}}\left[\frac{-i\pm 1}{\sqrt{F}} p^{}_x\right]
  e^{ -\frac{\sigma_0^2}{2} (p^{}_x - p^{}_0 )^2 - i ( p^{}_x - p^{}_0 ) x^{}_0}, \label{cpcm}
\end{equation}
where 
\begin{eqnarray}
{\cal N}^{}_{p^{}_x} &\equiv& 
 \left\{\left|D_{-\frac{1}{2}+\frac{iM^2}{2F}}\left[\frac{-i+1}{\sqrt{F}} p^{}_x\right] \right|^2+
 \left|D_{-\frac{1}{2}-\frac{iM^2}{2F}}\left[\frac{-i-1}{\sqrt{F}} p^{}_x\right] \right|^2
 \right\}^{-1}.
\end{eqnarray}

\begin{figure}[h] 
\includegraphics[width=5.7cm]{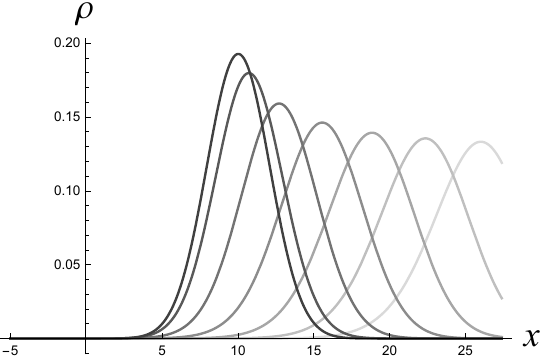}\hspace{.5cm}
\includegraphics[width=5.7cm]{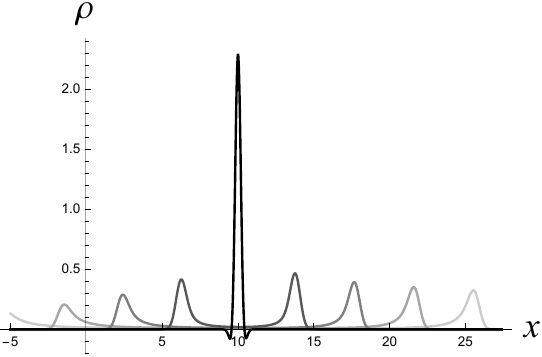} 
\includegraphics[width=5.7cm]{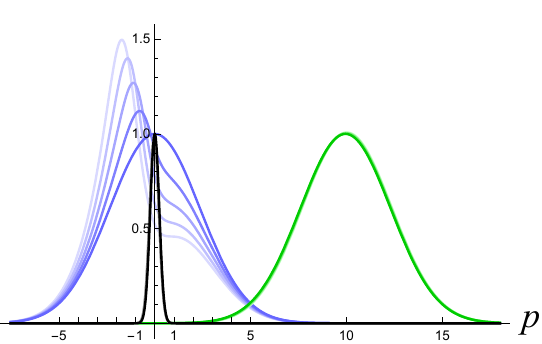} \hspace{.5cm}
\includegraphics[width=5.7cm]{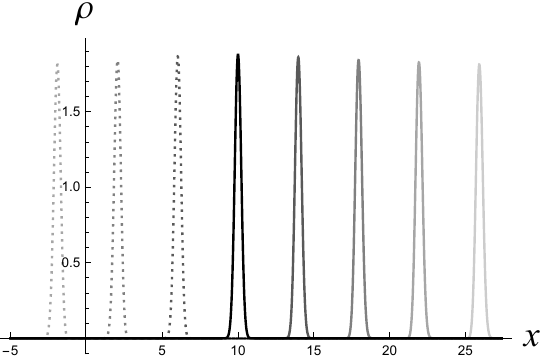}
\caption{Time evolution of the KG charge density $\rho(t,x)$ of wavefunction (\ref{PsiUAC}) with (\ref{psiDfn}) and (\ref{cpcm}). Here $F = 0.1$, $(\sigma^{}_0, \gamma^{}_0) = (3, 1)$ (upper-left), $(0.3, 1)$ (upper-right), and $(0.3, 10)$ (lower-right). Their momentum spectrum $\tilde{\rho} (t,p)\equiv |\psi^{}_{\bf p}(t)|^2$ with $\psi^{}_{\bf p}(t)$ in (\ref{psiDfn}) are shown in the lower-left plot, where the black, blue, and green curves represent the cases of $(\sigma^{}_0, \gamma^{}_0) = (3, 1)$, $(0.3, 1)$, and $(0.3, 10)$, respectively, with the peak-values of each case at $t=0$ normalized to 1. In the lower-left plot, only the blue curves have significant tails in the region of $p < -mc = -1$, corresponding to the double-peak structure in the upper-right plot for the KG charge density $\rho(t,x)$ with $(\sigma^{}_0, \gamma^{}_0) = (0.3, 1)$.
The solid curves from dark to light represent $\rho$ or $\tilde{\rho}$ at $t=0, 4, 8, 12, 16, \cdots$, while the dotted curves in the lower-right plot represent $\rho$ at $t=-4, -8$, and $-12$ from dark to\ light.}
\label{WavepacketInE}
\end{figure}

Since $A^0=0$ in this gauge, the charge density (\ref{ChargeDensityDef}) is simply $\rho = \frac{q\hbar}{mc^2} {\rm Re}\,\, i \Psi^* \partial_t \Psi$ here.
In Figure \ref{WavepacketInE}, we show three examples of the charge density $\rho$ of wavefunction (\ref{PsiUAC}) with (\ref{cpcm}), $F = 0.1$, and other parameter values similar to those in Figure \ref{GaussWavepacketFree}.
These examples correspond to the classical trajectories $\bar{x}(t) = c \sqrt{\alpha^{-2} + t^2}$ (the upper plots) and $\bar{x}(t)= c\sqrt{\alpha^{-2}+(t + t^{}_0)^2}-c\sqrt{\alpha^{-2}+t_0^2}+ c\alpha^{-1}$ (the lower-right plot) with $\alpha \equiv F/(mc)$ and $t^{}_0 \equiv p^{}_0/F$ for uniformly accelerated charges, and so $\bar{x}(0) = c\alpha^{-1} =10$ in all the three examples in Figure \ref{WavepacketInE}.

For a charged particle initially at rest, we find that $\rho$ can be close to a Gaussian function with the value of $\sigma^{}_0$ as small as 3 [Figure \ref{WavepacketInE} (upper-left)], {\it i.e.} with the initial width about $2\sigma^{}_0 = 6$ in the unit of the reduced Compton wavelength of the particle. This result is the same as we observed in Figure \ref{GaussWavepacketFree} (upper-left) for free particles.

\begin{figure}
\includegraphics[width=5.7cm]{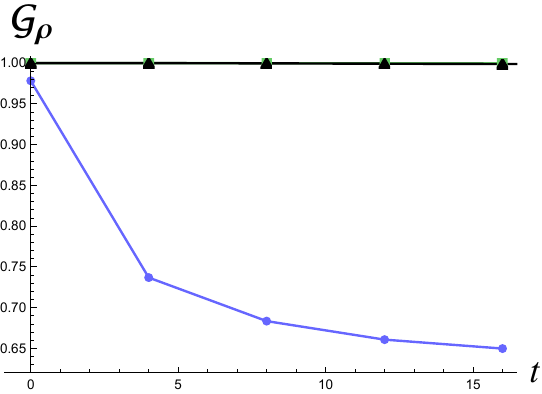}\hspace{.5cm}
\includegraphics[width=5.7cm]{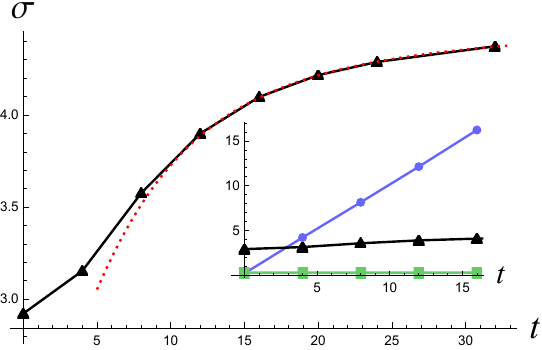}
\caption{(Left) ${\cal G}^{}_\rho$ of the KG charge densities in Figure \ref{WavepacketInE}. Here the black, blue, and green curves represent the cases of $(\sigma^{}_0, \gamma^{}_0)=(3,1)$, $(0.3, 1)$, and $(0.3, 10)$, respectively.
The imaginary part of ${\cal G}^{}_\rho$ is no greater than $O(10^{-4})$ in the worst case $(\sigma^{}_0, \gamma^{}_0)=(0.3, 1)$ (blue) and so negligible here. 
(Right) The half-widths $\sigma$ producing the values of ${\cal G}^{}_\rho$ in the left plot. The red dotted curve represents $f(t)=1.89 
+ 0.26 
\big( t/\sqrt{1+(\alpha t)^2} \, \big)$, $\alpha=0.1$, for comparison.}
\label{WavePacketinESimGauss}
\end{figure}

For $\sigma^{}_0 < 3$, as $\sigma^{}_0$ decreases, deflection of the charge density $\rho$ from a Gaussian function becomes more and more significant at $t=0$ (though $|\Psi(0,x)|^2$ is still Gaussian), and the non-Gaussianity of $\rho$ grows more and more quickly as $t$ increases [Figure \ref{WavePacketinESimGauss} (left).]  
In particular, as shown in Figure \ref{WavepacketInE} (upper-right), the KG charge density of a wavepacket with initial momentum $p^{}_0=0$ and initial width $\sigma^{}_0$ well below the Compton wavelength will quickly evolve to a double-peak structure, though the expectation value of position still moves around $\bar{x}(t)$ like a classical accelerated charge. This behavior is similar to those of free wavepackets with small initial widths in Figures \ref{WavePacketFree} and \ref{GaussWavepacketFree}. The significant tail of the corresponding momentum spectrum $\tilde{\rho}$ (blue curves) in the region of $p< -mc= -1$ in Figure \ref{WavepacketInE} (lower-left) is also similar to the momentum spectra of those wavepackets of free particles with initial widths well below the Compton wavelength in Figure \ref{WavePacketFreeP}.

When both the initial speed and the strength of the uniform electric field are small, the charge density $\rho$ of wavefunction (\ref{PsiUAC}) with initial width $2\sigma^{}_0$ above $O(\lambda^{}_C)$ will spread as $t$ increases, while the spreading rate decreases as the particle's classical speed $\dot{\bar{x}}(t)$ goes to the speed of light [Figure \ref{WavepacketInE} (upper-left)]. In Figure \ref{WavePacketinESimGauss} (right) we can see that the half-width $\sigma$ of the charge density $\rho$ in this case (black curve) evolves like $t/\bar{\gamma}(t)$ (red-dotted), rather than the proper time $\tau = \alpha^{-1} \sinh^{-1}\alpha t$, for sufficiently large $t$
\footnote{The worldline of our uniformly accelerated charge at proper acceleration $a$ in Figure \ref{WavepacketInE} (upper-left) and (upper-right) is $\bar{z}^{\mu}(\tau) = ( c\alpha^{-1} \sinh\alpha\tau, c \alpha^{-1} \cosh\alpha\tau,0,0 )$ parametrized by its proper time $\tau$, or $\bar{z}^{\mu}(t) = (ct, c \sqrt{\alpha^{-2}+t^2},0,0)$ parametrized by the Minkowski time $t$, with $\alpha\equiv a/c$ and $\bar{z}^\mu(0)=(0, c\alpha^{-1},0,0)$. It is straightforward to obtain the three velocity $\bar{v}^{i}\equiv d\bar{z}^i(t)/dt = (ct/\sqrt{\alpha^{-2}+t^2},0,0)$, and the Lorentz factor $\bar{\gamma}(t) = \big( 1- \frac{v^{}_i}{c}\frac{v_{}^i}{c} \big)^{-1/2}= \sqrt{ 1+(\alpha t)^2}$.}. 
This may be considered as a manifestation of length contraction, rather than time dilation, from the case of free particles at zero speed, whose half-width evolves like $t$ for sufficiently large $t$ (cf. Figure \ref{GaussWavepacketFreeSimGauss}). 

If the initial speed $v^{}_0$ of the charged particle at $t=0$ is close to the speed of light, then again, the minimal initial width for a long-lasting Gaussian charge density goes down to $O(1/\gamma^{}_0)$ of the Compton wavelength as shown in Figure \ref{WavepacketInE} (lower-right), where $\sigma^{}_0 = 3/\gamma_0 = 0.3$. This may also be considered as a manifestation of length contraction.

\section{Summary}
\label{summary}

We have demonstrated with selected examples that the charge density $\rho$ of the Klein-Gordon wavepacket of a relativistic particle in uniform motion could hopefully be approximated by the probability density of a Gaussian wavefunction in the Schr\"odinger representation of an effective theory with the single-particle interpretation, if the initial width of $\rho$ is above $O(\lambda^{}_C)/\gamma^{}_0$, where $\lambda^{}_C$ is the particle's Compton wavelength and $\gamma^{}_0$ is its Lorentz factor at the initial moment. For a particle with non-negative initial momentum $p^{}_0$, an initial width well above $\lambda^{}_C/\gamma^{}_0$ in position space corresponds to a momentum spectrum with a tail negligibly small in the region of $p < -mc$ in momentum space. For the wavepackets of free particles in uniform motion with all parameter values the same except the initial speed, Gaussian approximation could be good for a longer time for a particle at a higher initial speed in the laboratory (rest) frame.

The wavepackets of relativistic particles linearly accelerated in a uniform electric field show similar behaviors both in position space and momentum space. With an initial width of $\rho$ above $O(\lambda^{}_C)/\gamma^{}_0$, the spreading of the charge density of a uniformly accelerated Gaussian wavepacket gets frozen in the laboratory frame as the group velocity of the wavepacket approaches the speed of light. We find that this is a manifestation of length contraction, rather than time dilation.

Since our initial condition always produces wavepacket solutions spreading as time increases, the minimal initial width of $\rho$ we found in this paper actually implies that, to keep the single-particle interpretation and the Gaussianity, the charge density of a nearly Gaussian KG wavepacket at each moment $t$ can at most be squeezed to a width about $\lambda^{}_C/\bar{\gamma}(t)$.
This suggests that the UV cutoff of the electron-photon interaction could be about $\lambda^{}_C/\bar{\gamma}(t)$ in an effective theory quantized in Minkowski coordinates for {\it single electrons} interacting with EM fields \cite{LH23, Lin23}.\\

\noindent {\bf Acknowledgment}
The authors thank Prof. Jorma Louko for pointing out an issue in the earlier version of this paper.
SYL thanks Prof. Bei-Lok Hu for illuminating discussions. 
YCH and SYL are supported by the National Science and Technology Council of Taiwan under grant No. NSTC 112-2112-M-018-003 and in part by the National Center for Theoretical Sciences, Taiwan.

\begin{appendix}

\section{Phase and classical action}
\label{ApxPhiScl}

The classical action for a charged particle moving in EM fields in the Minkowski-time gauge reads \cite{Ro65, LH23, Lin23}
\begin{eqnarray}
  S^{}_{cl} &=& \int dt \left\{ -m c^2 \sqrt{1-\frac{1}{c^2}\frac{d z_i}{dt} \frac{dz^i}{dt}} + qc A_0(t,{\bf z}(t))
	 + q \frac{dz^i}{dt}A_i(t,{\bf z}(t)) \right\} \nonumber\\
	&\equiv& \int dt\, L^{}_{cl}, \label{clAction}
\end{eqnarray}
where $L^{}_{cl}$ is the classical Lagrangian.

\begin{figure} 
\includegraphics[width=3.9cm]{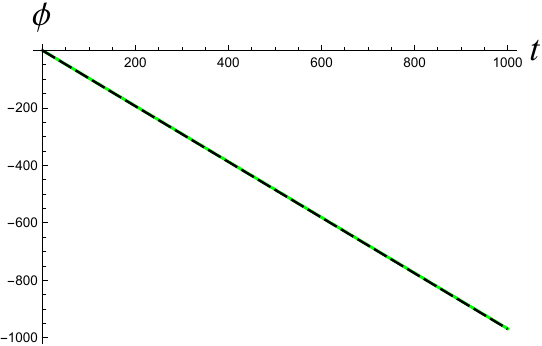} 
\includegraphics[width=3.9cm]{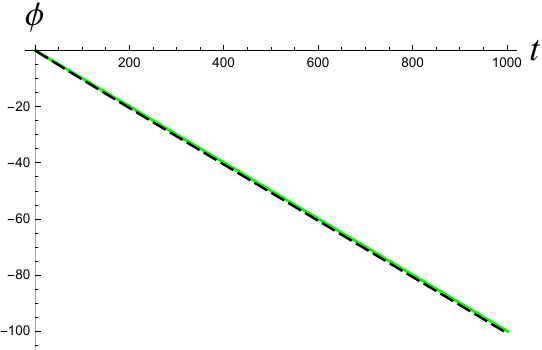}
\includegraphics[width=3.9cm]{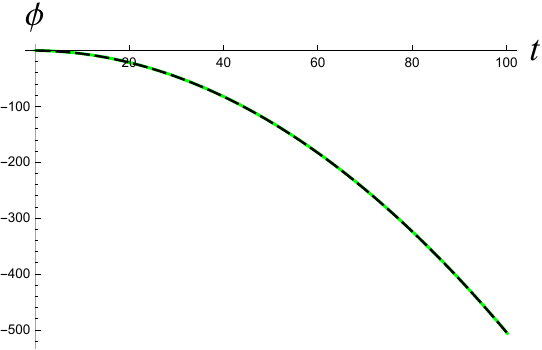}
\caption{ (Left) Time evolution of the phase $\phi(t,\bar{\bf x}(t))$ of wavefunction (\ref{PsiRelFree}) with $c\vartheta=100$ (black-dashed curve), compared with the classical action $S^{}_{cl}$ in (\ref{clActionFree}) and evaluated along the classical worldline of the particle $\bar{\bf x}(t)$ (green). Here $v=c/4$, and $c=\hbar=1$. 
(Middle) The phase $\phi(t,\bar{\bf x}(t))$ of (\ref{GaussianPsiFree}) with $\sigma^{}_0=0.3$ and $\gamma^{}_0 =10$ (black dashed), compared with $S_{cl}$ in (\ref{clActionFree}) (green). 
(Right) The phase $\phi(t,\bar{\bf x}(t))$ of wavefunction (\ref{PsiUAC}) with (\ref{psiDfn}) and (\ref{cpcm}) (black dashed), compared with $S_{cl}$ in (\ref{SclUAC}) (green). Here $(\sigma^{}_0,\gamma^{}_0) =(0.3, 10)$ and $F=0.1$, with other parameters the same as the left and middle plots.}
\label{PhaseWP}
\end{figure}

For a free particle in uniform motion at speed $v^{}_0$, one has $A^\mu=0$ and 
\begin{equation}
  S^{}_{cl} = \int dt (-m c^2) \sqrt{1-\frac{v_0^2}{c^2}}= -\frac{mc^2}{\gamma^{}_0} t, \label{clActionFree}
\end{equation}
where $\gamma^{}_0$ is the Lorentz factor of the particle motion. Compare (\ref{clActionFree}) and (\ref{clAction}), one obtains $L^{}_{cl}=-mc^2/\gamma^{}_0$ in this case.
Now writing (\ref{PsiRelFree}) for the free particle as 
\begin{equation}
  \Psi = R(t,{\bf x}) e^{i\phi(t,{\bf x})} \label{PsiPhase}
\end{equation}
with real functions representing magnitude $R$ and phase $\phi$. In Figure \ref{PhaseWP} (left), one can see that the phase of wavefunction (\ref{PsiRelFree}) evaluated along the particle's classical trajectory, namely, $\phi(t,\bar{\bf x}(t))$, behaves similarly to the particle's classical action $S_{cl}$ in (\ref{clActionFree}) for sufficiently large $t$, as expected.   

Indeed, when $t$ is sufficiently large and $x = \bar{x}(t) = x_0 + v_0 t$, one has $F(t,\bar{x}(t)) \approx ic\sqrt{1-(v_0/c)^2}\,t$ in (\ref{PsiRelFree}), and the phase of $\Psi(t,\bar{x}(t),0,0)$ behaves like $- mc^2 \sqrt{1-(v_0/c)^2}\,t/\hbar 
= S^{}_{cl}/\hbar$ 
according to the asymptotic behavior of the modified Bessel functions \cite{GR07}, from which one can also see $\phi(t,\bar{\bf x}(t))-S^{}_{cl}(t) \to -\pi/4$ as $t\to \infty$. 

As shown in Figure \ref{PhaseWP} (middle), the phase of wavefunction (\ref{GaussianPsiFree}) for a free particle behaves like the classical action (\ref{clActionFree}), too, when $t$ is sufficiently large.

For a charged particle moving in a uniform electric field considered in Section \ref {KGuEF}, we have $A^\mu = (0, -{\cal E} t,0,0)$, $\frac{d}{dt}\bar{x} = c\alpha (t+t^{}_0)/\sqrt{1+ [\alpha(t+t^{}_0)]^2}$, and $\frac{d}{dt}\bar{y}=\frac{d}{dt}\bar{z}=0$, so (\ref{clAction}) reads
\begin{equation}
  S^{}_{cl} = - mc^2 \int_0^t d\tilde{t} \,
	  \frac{1 + \alpha^2\tilde{t}\left(\tilde{t}+t^{}_0\right)}{\sqrt{1+ \alpha^2\left(\tilde{t}+t^{}_0\right)^2}} \label{SclUAC}
\end{equation}
When $\alpha(t+t^{}_0)$ is sufficiently large, $S^{}_{cl} \sim -mc^2 \int^t d\tilde{t} \alpha \tilde{t} = -mc^2\alpha t^2/2 = -cF t^2/2$.
In figure \ref{PhaseWP} (right), one can see that the phase $\phi(t,{\bf x}(t))$ of wavefunction (\ref{PsiUAC}) with (\ref{psiDfn}) and (\ref{cpcm}) also behaves like $S^{}_{cl}$ in (\ref{SclUAC}) for sufficiently large $t$.

\end{appendix}

\end{document}